\documentclass[3p]{elsarticle}

\usepackage{graphicx}
\usepackage[colorlinks=true]{hyperref}
\usepackage{amssymb}
\usepackage{siunitx}
\usepackage{upgreek} % For for non-italic greek letters
\usepackage{marginnote}
\usepackage{color} % For using colors in the correction of the manuscript
\definecolor{blue}{rgb}{0,0,0.8}

\definecolor{red}{rgb}{0.8,0,0}

\definecolor{green}{rgb}{0,0.4,0}

\definecolor{violet}{rgb}{0.8,0.2,0.8}
\definecolor{darkgreen}{RGB}{9, 90, 6}

\usepackage[symbol]{footmisc}

\usepackage[normalem]{ulem}
\usepackage{multirow}
\usepackage{multicol}
\usepackage{nonfloat}

\definecolor{darkPurple}{RGB}{150,0,150}
\definecolor{lightBrown}{rgb}{0.7,0.7,0.5}

\usepackage{ifthen}
\newboolean{clean}
\setboolean{clean}{false} 

\ifthenelse{\boolean{clean}}
{
    \newcommand\commentOld[1]{\bigskip}
    
}
{

    \newcommand\commentOld[1]{{\color{lightBrown} #1}}
    
}

\usepackage{amsmath} 
\usepackage{lineno}

\newcommand{\n}{\hat{\mathbf{n}}}
\newcommand{\g}{\hat{\mathbf{g}}}

\usepackage[symbol]{footmisc}
\biboptions{authoryear}
\journal{NeuroImage} %Journal Name
\makeatletter
\def\ps@pprintTitle{%
  \let\@oddhead\@empty
  \let\@evenhead\@empty
  \let\@oddfoot\@empty
  \let\@evenfoot\@oddfoot
}
\makeatother

\begin{document}
\begin{frontmatter}
\title{Assessment of Precision and Accuracy of Brain White Matter Microstructure using Combined Diffusion MRI and Relaxometry}

\author[1,2]{Santiago Coelho\corref{cor1}}
\author[1,2]{Ying Liao\corref{cor1}}
\author[3]{Filip Szczepankiewicz}
\author[1,2]{Jelle Veraart}
\author[1,2]{Sohae Chung}
\author[1,2]{Yvonne W. Lui}
\author[1,2]{Dmitry S. Novikov}
\author[1,2]{Els Fieremans}
\address[1]{Bernard and Irene Schwartz Center for Biomedical Imaging, Department of Radiology, New York University Grossman School of Medicine, New York, NY, USA}
\address[2]{Center for Advanced Imaging Innovation and Research (CAI$^2$R), Department of Radiology, New York University Grossman School of Medicine, New York, NY, USA}
\address[3]{Medical Radiation Physics, Clinical Sciences Lund, Lund University, Lund, Sweden}
\cortext[cor1]{Ying Liao and Santiago Coelho equally contributed to the manuscript}

%=================================================
%============= START OF ABSTRACT =================
%=================================================

\begin{abstract}
Joint modeling of diffusion and relaxation has seen growing interest due to its potential to provide complementary information about tissue microstructure. For brain white matter, we designed an optimal diffusion-relaxometry MRI protocol that samples multiple b-values, B-tensor shapes, and echo times (TE). This variable-TE protocol (27 min) has as subsets a fixed-TE protocol (15 min) and a 2-shell dMRI protocol (7 min), both characterizing diffusion only. We assessed the sensitivity, specificity and reproducibility of these protocols with synthetic experiments and in six healthy volunteers. Compared with the fixed-TE protocol, the variable-TE protocol enables estimation of free water fractions while also capturing compartmental $T_2$ relaxation times. Jointly measuring diffusion and relaxation offers increased sensitivity and specificity to microstructure parameters in brain white matter with voxelwise coefficients of variation below 10\%. 

\end{abstract}


\end{frontmatter}

\begin{multicols}{2}

\section{Introduction}
\noindent

Multi-compartment models are widely used in MRI studies of biological tissues. Compartmentalization allows for dividing a complex heterogeneous structure into simpler, homogeneous units. In the brain white matter (WM), there are at least four major compartments: intra-axonal space (IAS), myelin, extra-axonal space (EAS) and free water (FW). Each compartment is distinguished by unique diffusion profiles \citep{novikov2019} and $T_2$ relaxation times \citep{veraart2018te}. For instance, the $T_2$ relaxation time of myelin is 10--40 ms \citep{mackay1994vivo}, while $T_2$ of the white matter tissue is 70--100 ms \citep{stanisz2005t1}, and $T_2$ of cerebrospinal fluid (CSF) is 500--2100 ms \citep{piechnik2009functional}. Therefore, the signal of myelin water decays much faster with the echo time (TE) than the other compartments and is typically not observable for clinical diffusion MRI (dMRI) protocols. However, modeling either relaxation or diffusion properties independently presents challenges in parameter estimation, even for a simple two-compartment model \citep{jelescu2016degeneracy}. Combining diffusion and relaxation holds tremendous promise, as these two modalities provide complementary information for every compartment \citep{veraart2018te, lampinen2023probing, tax2021measuring}.

\begin{figure*}[b]
\centering
\includegraphics[width=\textwidth]{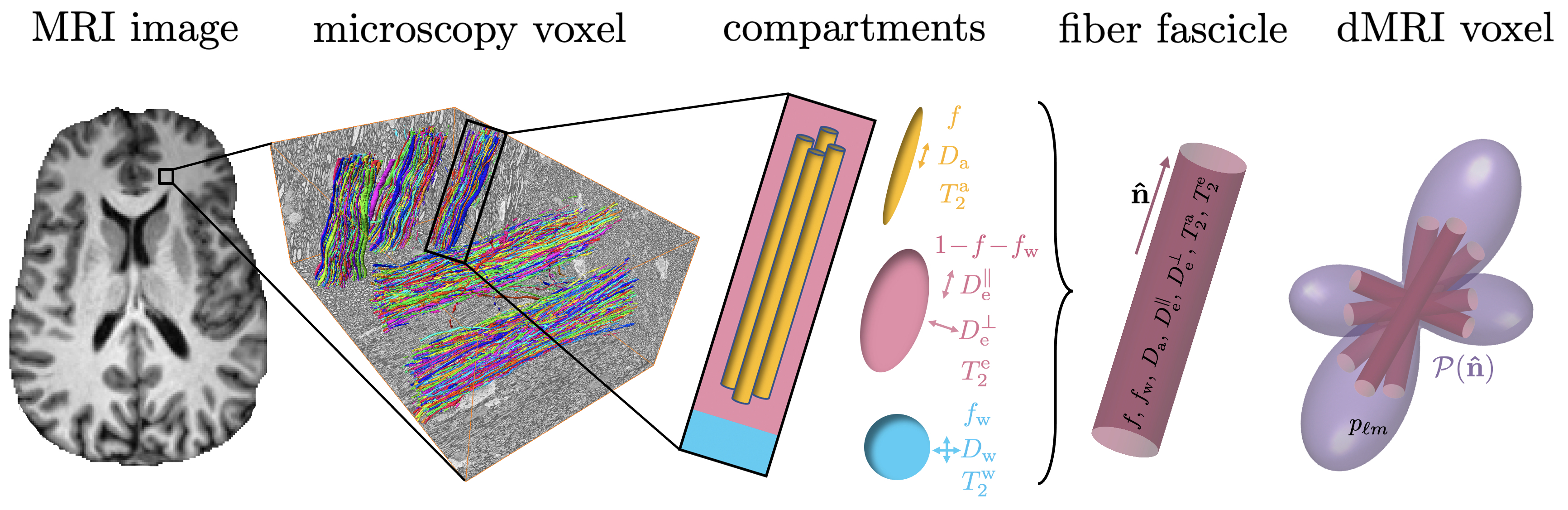}
\caption[caption FIG]{Diagram of the Standard Model of diffusion-relaxation in white matter. In an MRI voxel, multiple elementary fiber fascicles coexist (whose MRI signal yields the so-called fiber response or \textit{kernel}). Fiber fascicles contain multiple compartments that are characterized by their diffusivities, relaxivities, and water fractions. An MRI voxel is thus modeled as a collection of such fascicles oriented following an arbitrary fODF $\mathcal{P}(\hat{n})$.}
\label{fig:SM_diagram}
\end{figure*}

The FW compartment, though small in most WM voxels, gains significance at WM-CSF boundaries or in cases of edema. It is characterized by isotropic diffusivity and considerably longer $T_2$ relaxation times compared to the IAS or EAS. Quantifying FW fractions is of great interest for clinical applications, such as inflammation \citep{jelescu2023sensitivity}, schizophrenia \citep{pasternak2012excessive} and Parkinson's disease \citep{planetta2016free}. Techniques have been proposed to model the FW compartment in addition to diffusion tensor imaging (DTI) \citep{pasternak2009free}, or to diffusion kurtosis imaging (DKI) \citep{collier2018diffusion}. Yet, the inverse problem of estimating FW fractions solely from diffusion data is still ill-conditioned \citep{golub2021free}. Biophysical models of diffusion have routinely omitted FW to avoid these instabilities. To resolve this, multidimensional diffusion weightings have been demonstrated to facilitate solving FW fractions in a three-compartment system analytically \citep{reisert2019unique}. By utilizing the differences in diffusion and $T_2$ relaxation times between FW and other compartments, the combined diffusion-relaxometry approach may enhance the sensitivity of multi-compartment models to FW, thereby improving the overall accuracy of parameter estimation \citep{anania2022improved}.

The Standard Model (SM) of WM was proposed as an overarching framework \citep{novikov2019} that encompasses many previously proposed multi-compartment dMRI models \citep{kroenke2004,jespersen2007,assaf2004,alexander2010,fieremans2011,zhang2012,kaden2016,reisert2017,assaf2008,jespersen2010,sotiropoulos2012,reisert2017,novikov2018rotationally}. WM voxels are composed of a collection of fiber fascicles, which serve as the elementary building block for generating diffusion signals. Each fiber fascicle is assumed to consist of several non-exchanging compartments representing the IAS, EAS and FW. The overall dMRI signal is the weighted sum of the contributions from all bundles inside a voxel, each oriented according to the fiber orientation distribution function (fODF), see Fig.~\ref{fig:SM_diagram}. 
The SM can be extended onto the relaxometry by assigning a $T_2$ relaxation time to each compartment, and varying the TE via a joint diffusion-relaxation protocol \citep{veraart2018te, tax2021measuring,lampinen2019searching}.

An extended MRI protocol for diffusion-relaxation beyond conventional clinical protocols is necessary to overcome the degeneracy in parameter estimation. For instance, varying TE alters the relative weights between compartments with different $T_2$ values \citep{benjamini2016use, veraart2018te, ning2019joint,lampinen2019searching}. In addition, strong diffusion-weightings make dMRI measurements more sensitive to tissue microstructure \citep{novikov2019,jespersen2010neurite}. Multidimensional dMRI encodes diffusion along multiple directions simultaneously in the same diffusion weighting, thereby probing the response to a $3\times3$ $\textit{\textbf{B}}$-tensor with $\mathrm{rank\,} \textit{\textbf{B}} > 1$ \citep{mitra1995multiple,lasivc2014microanisotropy,topgaard2017multidimensional}, and hence provides richer characterization of tissue microstructure than conventional dMRI encodings along a single direction \citep{coelho2019resolving}.

Our aim has been to optimize a comprehensive diffusion-relaxation MRI protocol within 30 minutes, combining multiple TE, high b-values and multidimensional diffusion encodings, and evaluate its sensitivity, specificity, and reproducibility. This work builds further upon the protocol optimization by \cite{coelho2022reproducibility}, via incorporating varying TE. \cite{lampinen2020towards} previously designed a 15-minute acquisition protocol for diffusion-relaxometry by optimizing for the Cram\'er-Rao Lower Bound (CRLB) of the SM parameters. While the CRLB represents a theoretical lower bound of the variance for an unbiased estimator, most SM estimators are biased in conventional datasets. Machine learning (ML)-based estimators are not exempt from biases as they tend to push estimates towards the prior mean due to the limited obtainable information and SNR \citep{coelho2021we}, but they have shown to reduce overall estimation error significantly \citep{coelho2022reproducibility}. Thus, we adopted the root-mean-squared-error (RMSE) of an ML-based estimator \citep{reisert2017} as the objective function for our protocol design.

In the following sections, we will first explain the SM for multidimensional dMRI at varying TE. Next, we will present our framework of protocol optimization and the optimized variable-TE protocol (27 min). Then, we will compare the performance of the variable-TE protocol we proposed with its two subsets, a fixed-TE multidimensional diffusion protocol (15 min) and a 2-shell dMRI protocol (7 min) that is commonly used for DKI \citep{jensen2005diffusional}. We will assess the sensitivity and specificity of each protocol using the recently proposed Sensitivity-Specificity Matrix (SSM) \citep{liao2023mapping}, and we will evaluate the reproducibility of each using the coefficient of variation (COV) from the scan and rescan of six healthy volunteers. Furthermore, we will assess the feasibility of estimating FW fractions using a 2-shell dMRI protocol along with $\mathrm{b_0}$ measurements at multiple TEs (8 min). 

\begin{figure*}[htbp]
\centering
\includegraphics[width=\textwidth]{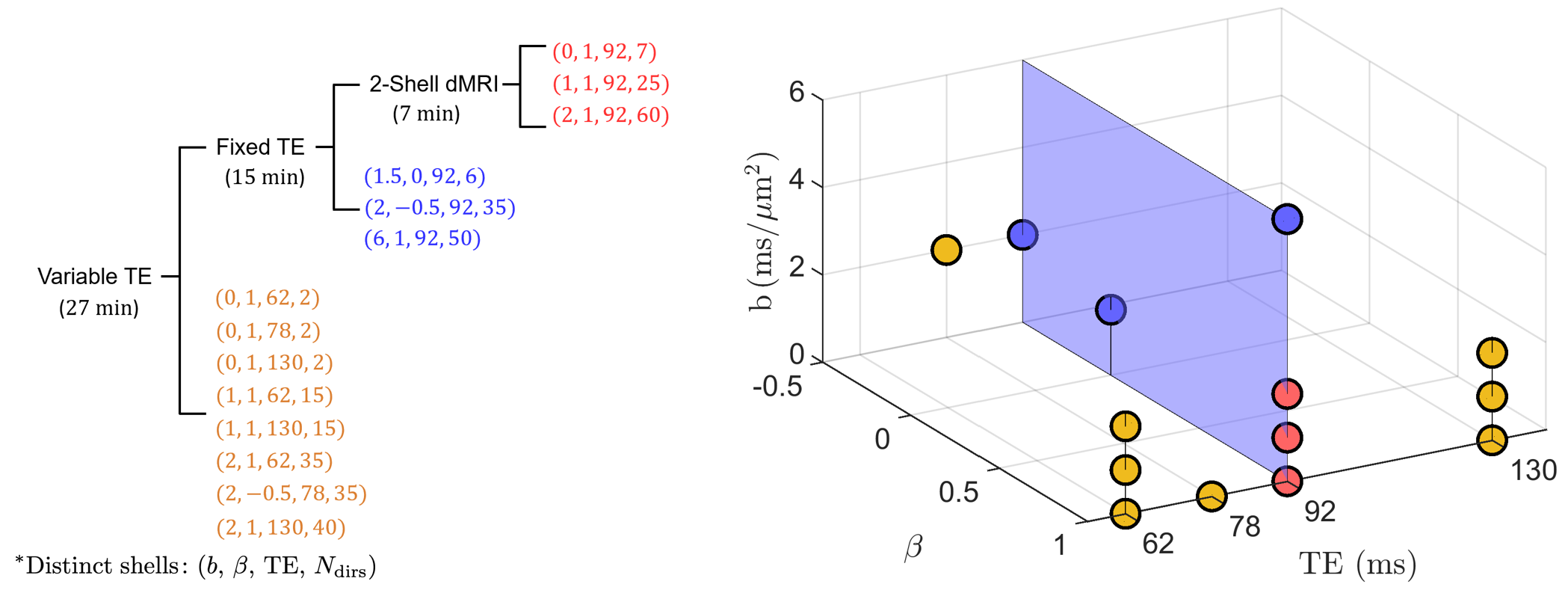}
\caption[caption FIG]{Hierarchy of the optimized MRI acquisition protocols. Each distinct shell is represented in the form of $(b, \beta, \mathrm{TE}, N_{\text{dirs}})$, where $b$ denotes b-value, $\beta$ B-tensor shape, $\mathrm{TE}$ echo time, $N_{\text{dirs}}$ number of gradient directions. The broader fixed-TE dMRI protocol encompasses the 2-shell dMRI protocol, as well as a low-b PTE and a STE shell, in addition to a high-b LTE shell. Building upon the fixed-TE protocol, the variable-TE protocol introduces shells with TE values of 62 ms, 78 ms, and 130 ms. A 3D plot visualizes these protocols within the acquisition parameter space, accompanied by a tree structure for clear representation of their relationship. In this plot, each dot is color-coded to represent a distinct shell. Notably, the 2-shell dMRI and fixed-TE protocol are distributed on a blue plane with $\mathrm{TE} = 92$ ms.}
\label{fig:fig1_protocol}
\end{figure*}

\section{Theory}
\subsection{$T_2$ relaxometry and multidimensional dMRI}
\noindent
$T_2$ relaxometry has been widely used in conjunction with multi-compartment models to study volume fractions of myelin, axons, and FW \citep{mackay1994vivo,whittall1997vivo}. 
In a voxel with multiple water pools, each with its own $T_2$ relaxation, the overall signal fraction is:
\begin{equation}\label{eq_T2_compartments}
    S(\mathrm{TE}) = s_0 \sum_i f_i e^{-\mathrm{TE}/T_2^i},\quad \sum_i f_i = 1,
\end{equation}
where $s_0$ is the proton density, each $f_{i}\geq0$. In the continuum limit, Eq.~(\ref{eq_T2_compartments}) becomes a Laplace transform of the density spectrum of $T_2$ values. 
Thus, sampling MRI data at multiple TE can, in principle, enable the quantification of relaxation properties from different water pools. Although this framework has been widely applied, it is ill-conditioned, just like the inverse Laplace transform \citep{epstein2008bad}. 
Small fluctuations in the measured signals can lead to a wide range of reasonable outputs, making it unstable for the signal-to-noise ratio (SNR) of conventional datasets.

dMRI provides a similar, yet complementary, tissue probe to relaxometry. Rather than generating signal decay due to the relaxation of the transverse magnetization, these measurements encode information about the random motion of water molecules. Conventional dMRI uses linear tensor encoding (LTE) to probe diffusion along a single direction within a given diffusion weighting. To probe tissue anisotropy, these measurements are performed along multiple directions by rotating the diffusion encodings. In contrast, multidimensional dMRI \citep{westin2016q,topgaard2017multidimensional} simultaneously encodes diffusion along multiple directions within one diffusion weighting. This generalizes the scalar b-value to a B-tensor:
\begin{equation}\label{eq_qt}
\begin{aligned}
     B_{ij} &= \int_{0}^{\mathrm{TE}}q_{i}(t)q_{j}(t)dt,\;\; 
     q_{i}(t)=\int_{0}^{t}g_{i}(t')dt',
\end{aligned}
\end{equation}
where $g(t)$ is the effective diffusion gradient. The trace $b=\mathrm{tr}\, \boldsymbol{B}$ is the conventional $b$-value. These additional experimental settings provide complementary information to LTE and their combination better informs biophysical modeling \citep{coelho2019resolving,reisert2019unique,lampinen2019searching,lampinen2023probing}. 

In this work, we focus on axially symmetric B-tensors, which can be represented as \citep{topgaard2017multidimensional}
\begin{equation}\label{eq_tensor_shape}
    B_{ij}(b,\beta,\g) = b\left(\beta\, g_{i}g_{j} + \frac{1-\beta}{3}\,\delta_{ij}\right),
\end{equation}
where $\g$ is the gradient direction (the symmetry axis) and $\beta$ is the shape parameter. In particular, $\beta=1$ for linear tensor encoding (one nonzero eigenvalue, LTE), $\beta=-\frac{1}{2}$ for planar tensor encoding (two nonzero eigenvalues, PTE) and $\beta=0$ for spherical tensor encoding (three identical nonzero eigenvalues, STE). In other words, the rank of the B-tensor determines the number of dimensions being probed by the diffusion gradient.

An axially symmetric diffusion tensor 
$D_{ij} = D^\bot \delta_{ij} + (D^{\parallel}-D^\bot) n_i n_j$ is characterized by its parallel and perpendicular diffusivities $D^{\parallel}$ and $D^{\bot}$, and by a unit vector $\n$ defining its principal axis. When probing axially symmetric Gaussian compartments with axially symmetric B-tensors (\ref{eq_tensor_shape}), the diffusion attenuation (assuming the unattenuated signal is normalized to $1$) becomes
\begin{equation}\label{eq_BD}
\begin{aligned}
    S(B) &= e^{-\sum_{ij} B_{ij} D_{ij}}\\
    &= e^{-b(D^{\parallel}-D^{\bot})\left(\beta(\xi^2-\tiny{\frac{1}{3}}\normalsize)+\tiny{\frac{1}{3}}\normalsize\right)-bD^{\bot}},
   % \sum_{ij} B_{ij} D_{ij}= -b(D^{{\parallel }}-D^{\bot})\left(\beta(\xi^2-\tfrac{1}{3})+\tfrac{1}{3}\right)\\
    % -bD^{{\bot}},
    \end{aligned}
\end{equation}
where $\xi =\g\mathrm{\cdot }\n$. 

\subsection{SM for diffusion-relaxometry}
\noindent
In the SM framework, WM voxels are modeled as a collection of fiber segments (fascicles), elementary bundles of aligned axons, see Fig.~\ref{fig:SM_diagram}. The fiber fascicle contains three non-exchanging compartments: IAS, EAS and FW. Myelin is not visible at the TE used in clinical dMRI.

The IAS compartment models axons as zero-radius cylinders (sticks) with volume fraction $f$, relaxation time $T_2^\mathrm{a}$ and diffusivity $D_\mathrm{a}$ along the axon direction. The EAS compartment represents hindered diffusion with relaxation time $T_2^\mathrm{e}$ and diffusivities parallel and perpendicular to axons $D_\mathrm{e}^{\parallel}$ and $D_\mathrm{e}^{\bot}$, respectively. The FW compartment with fraction $f_\mathrm{w}$ and relaxation time $T_2^\mathrm{w}$ has an isotropic diffusivity $D_\mathrm{w}=3$ \unit{\mu m^2/ms} (water diffusivity at body temperature). 

At clinically relevant experimental settings, tissue compartments behave approximately Gaussian. Thus, the signal response of a single fiber fascicle --- the kernel --- is a sum of signals from three Gaussian compartments weighted by their fractions and $T_2$ values. For an arbitrary diffusion-weighting and TE, the kernel is
\begin{equation}\label{eq_kernel}
\begin{aligned}
    \mathcal{K}(&b,\beta,\xi,\mathrm{TE})=f\,e^{-bD_\mathrm{a}\left(\beta(\xi^2-\tfrac{1}{3})+\tfrac{1}{3}\right)-\mathrm{TE}/T_2^\mathrm{a}}\\
    &+f_\mathrm{e}\,e^{
    -b(D^{{\parallel }}_\mathrm{e}-D^{{\bot }}_\mathrm{e})\left(\beta(\xi^2-\tfrac{1}{3})+\tfrac{1}{3}\right)-bD^\bot_\mathrm{e}-\mathrm{TE}/T_2^\mathrm{e}}\\
    &+f_\mathrm{w}\,e^{-bD_\mathrm{w}-\mathrm{TE}/T_2^\mathrm{w}},\\
\end{aligned}
\end{equation}
where $f_\mathrm{e}=1-f-f_{\mathrm{w}}$ and $\xi =\g\mathrm{\cdot }\n$ is the scalar product between the symmetry axis of the kernel, $\n$, and the symmetry axis of the B-tensor, $\g$.

In an imaging voxel, fiber segments are typically not aligned but are oriented according to a probability distribution $\mathcal{P}(\n)$, the fODF. Therefore, the SM signal becomes the convolution of the kernel response and the fODF:
\begin{equation} \label{sph-conv}
S(b,\beta,\g,\mathrm{TE})=s_{0}\int_{\mathbb{S}^2}d\n\,\mathcal{P}(\n)\,\mathcal{K}(b,\beta,\g\mathrm{\cdot }\n,\mathrm{TE}),
\end{equation}
where $s_0$ is the proton density and $\int_{\mathbb{S}^2} d\n \mathcal{P}(\n)=1$. 
Eq.~(\ref{sph-conv}) has a general form of the spherical convolution that has been widely used in dMRI
\citep{healy1998,tournier2004,anderson2005,tournier2007}, where the assumptions on the physics of diffusion and relaxation determine the specific functional form (\ref{eq_kernel}) of the kernel. 

\subsection{Factorization in the spherical harmonics basis}
\noindent
The spherical harmonics (SH) basis is a natural choice for functions on a sphere. Here, we can write the fODF as:
\begin{equation}
\mathcal{P}(\n)\approx1+\sum_{\ell=2,4,...}^{\ell_{max}}\sum_{m=-l}^{\ell}p_{\ell m}Y_{\ell m}(\n),
\end{equation}
where $Y_{\ell m}(\n)$ are the SH basis functions and $p_{\ell m}$ are the SH coefficients. Likewise, the SM signal can be written in the SH basis, where the convolution becomes a product:
\begin{equation}
\begin{aligned}
    S(b,\beta,\g,\mathrm{TE}) &= \sum_{\ell=0,2,...}^{\ell_{max}}\sum_{m=-l}^{\ell}S_{\ell m}(b,\beta,\mathrm{TE})Y_{\ell m}(\g),\\
    S_{\ell m}(b,\beta,\mathrm{TE})&=s_{0}\,\mathcal{K}_{\ell}(b,\beta,\mathrm{TE})\,p_{\ell m}
\end{aligned}
\end{equation}
where 
\begin{equation}
    \mathcal{K}_{\ell}(b,\beta,\mathrm{TE})=\int_0^{1}d\xi\,\mathcal{K}(b,\beta,\xi,\mathrm{TE})\,P_{\ell}(\xi)
\end{equation}
are the projections of the SM kernel onto the Legendre polynomials $P_{\ell}(\xi)$ (nonzero $m$ projections are zero due to the axial symmetry of the kernel).

To remove the dependence on the choice of physical basis in three-dimensional space, we define the rotational invariants of the signal and fODF \citep{novikov2019}:
\begin{equation}
\begin{aligned}
      S^{2}_{\ell}(b,\beta,\mathrm{TE}) &= \frac{1}{4\pi(2\ell+1)}\sum_{m=-\ell}^{\ell}\lvert S_{\ell m}(b,\beta,\mathrm{TE})\rvert^2,\\  
      p_{\ell}^2 &= \frac{1}{4\pi(2\ell+1)}\sum_{m=-\ell}^{\ell}\lvert p_{\ell m}\rvert^2\,.
\end{aligned}
\end{equation}
The fODF rotational invariants are normalized such that $p_0=1$ and $0\leq p_\ell\leq 1$. Here, $p_\ell$ can become a metric of fODF anisotropy for any $l>0$, and we typically choose $p_2$ because it has higher SNR than the other high-order terms.

Thus, we can represent all rotationally invariant information in the SM signal as:
\begin{equation}
    S_{\ell}(b,\beta,\mathrm{TE})=s_{0}\,p_{\ell}\,\mathcal{K}_{\ell}(b,\beta,\mathrm{TE}),\;\ell=0,2,...\,.
\end{equation}
Note that rotational invariants of odd $\ell$ are zero due to time-reversal symmetry of the Brownian motion. Through this transform, we effectively compress the SM signal $S(b,\beta,\g,\mathrm{TE})$ to a few rotational invariants without loss of information in terms of estimating the kernel parameters.%, since the kernel is also rotationally invariant.

\begin{figure*}[b]
\centering
\includegraphics[width=\textwidth]{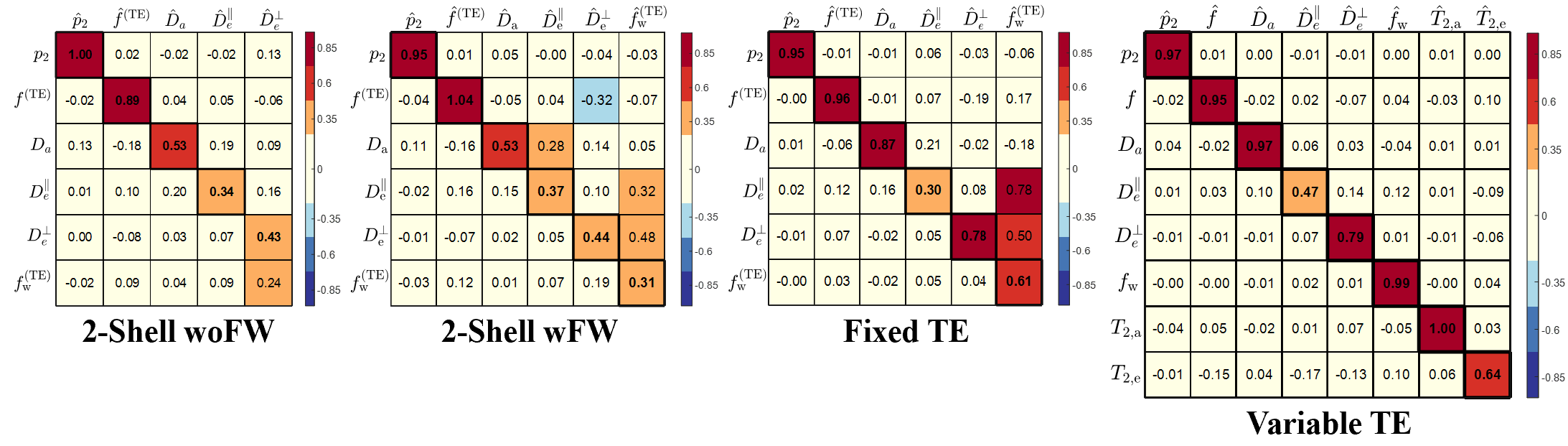}
\caption[caption FIG]{Sensitivity-specificity matrix of the 3 protocols. Diagonal elements measures the sensitivity of the estimation, while off-diagonal elements quantify (lack of) specificity. According to the SSM, the fixed-TE protocol significantly improves the estimation of $D_\mathrm{a}$ and $D_\mathrm{e}^{\bot}$, and the variable-TE protocol can further accurately estimate FW fractions as well as capture compartment $T_2$ values.}
\label{fig:fig2_ssm}
\end{figure*}

\section{Methods}
\subsection{MRI protocol optimization}
\noindent
The MRI acquisition protocol determines the amount of microstructure information encoded into the MRI signals. Previous studies have suggested high b-values \citep{novikov2019,jespersen2010neurite}, multidimensional dMRI \citep{mitra1995multiple,lasivc2014microanisotropy,topgaard2017multidimensional} and varying TE \citep{benjamini2016use, veraart2018te, ning2019joint,lampinen2019searching} can provide additional information to conventional 2-shell dMRI protocols. Here, we design our protocol by incorporating all of these features. We optimize our protocol based on the RMSE of an ML-based estimator (see \citep{coelho2022reproducibility} and Section \ref{SMI} for more details)
\begin{align}\label{eq_loss_fun}
    F(b,\beta,\mathrm{TE},N_\mathrm{dirs}) &= \sum_{i}\frac{\epsilon_{\theta_{i}}}{w_{i}},\;\mathrm{where}\\
    \epsilon_{\theta_{i}} &= \sqrt{\tfrac1{N_\mathrm{train}}\sum_{j}^{N_{\mathrm{train}}}(\hat{\theta}_{i;j} - \theta_{i;j})^2}
\end{align}
 $\epsilon_{\theta_{i}}$ is the RMSE for a given SM parameter $\theta_{i}$, $N_{\mathrm{train}}$ is the number of training samples for the ML-based estimator. Each RMSE term $\epsilon_{\theta_{i}}$ is normalized by the range of its parameter $w_i$. 

The full parameter set is defined as $\theta=\{p_2,f,D_\mathrm{a},D_\mathrm{e}^{\parallel},D_\mathrm{e}^{\bot},f_{\mathrm{w}},T_{2}^{\mathrm{a}},T_{2}^{\mathrm{e}}\}$. The $T_2$ value of the FW compartment is much longer than that of the IAS and EAS compartment and is set to 500 ms in our model. The objective function $\boldsymbol{F}$ is minimized by finding the optimal combinations of shells with $(b,\beta,\mathrm{TE},N_{\mathrm{dirs}})$, standing for b-value, B-tensor shape, echo time and number of distinct gradient directions, respectively. The individual RMSE was computed analytically as a function of the shells, SNR and the moments of the training set distribution as in \citep{coelho2021we}. We minimized Eq.~(\ref{eq_loss_fun}) using the Self Organizing Migrating Algorithm \citep{zelinka2016soma}.

The total acquisition time was constrained within 27 minutes. The maximum b-value was set to 10 \unit{ms/\mu m^2}. Given the hardware limitations such as maximum gradient amplitude and slew rate, we calculated the minimum encoding time for a given $\{b,\beta\}$ combination \citep{sjolund2015constrained,lampinen2020towards}. Each shell was assumed to have diffusion encodings along gradient directions uniformly distributed on a sphere for optimal acquisition. Furthermore, we fixed conventional multi-shells (2-shell dMRI protocol) into our protocol together with the protocol used by \cite{coelho2022reproducibility} (fixed-TE protocol), see Fig. \ref{fig:fig1_protocol}, to enable future comparisons.

\subsection{Parameter estimation} \label{SMI}
\noindent
Conventional maximum likelihood estimation (MLE) is unstable for SM parameter estimation at realistic SNR levels or with limited data. %due to the degeneracy in parameter likelihood landscape \citep{jelescu2016degeneracy, novikov2018rotationally}. 
\cite{reisert2017} first proposed the use of ML-based algorithms to estimate SM parameters, which have shown lower estimation error than conventional MLE. The ML-based estimator learns the mapping from synthetic MRI signals to model parameters from synthetic data generated from the model. %\cite{reisert2017} also demonstrated that a cubic polynomial regression is sufficient to interpolate this mapping at a realistic SNR:
Such mapping can be written as:
\begin{equation}
    \hat{\theta}=\sum_{j_1+j_2+...j_N\leq W}a_{j_1,j_2,...,j_N}y_{1}^{j_1}y_2^{j_2}...y_N^{j_N} \,,
\end{equation}
where $\hat{\theta}$ is the ML estimator, ${y_i}_{i=1}^{N}$ are the noisy measurements (rotational invariants $S_l(b,\beta,\mathrm{TE})$ in our case), $W$ is the degree of the polynomial regression, and $a_{j_1,j_2,...,j_N}$ are the regression coefficients computed during training, which are a function of SNR. \cite{coelho2021we} demonstrated that a cubic polynomial regression is sufficient to approximate this mapping at realistic SNR levels and that more complex regressions, e.g. fully connected neural networks or random forests, lead to identical RMSE values. After generating signals based on Eqs.~(\ref{eq_kernel})(\ref{sph-conv}), Gaussian noise was added to the signal, as we assumed the Rician floor was minimized by denoising the complex-valued data \citep{lemberskiy2019achieving}. %removed in the preprocessing \citep{koay2006analytically}. 
During the inference, we selected the corresponding regression coefficients based on the estimated SNR  \citep{veraart2016diffusion}. The code for the ML-based estimation is now publicly available in the Standard Model Imaging (SMI) toolbox (\url{https://github.com/NYU-DiffusionMRI/SMI}).

The performance of SMI, like other ML algorithms, is influenced by the nature of its training data. This impact is mediated by the acquired measurements and their SNR. In scenarios where comprehensive protocols are employed, the algorithm's performance is less susceptible to variations in training data. Conversely, in cases with limited protocols containing less information, the performance is more sensitive to changes in the training data. Here, we used a uniform distribution for training to minimize the impact of prior information, with lower bounds [0.05, 0.05, 1, 1, 0.1, 0, 50, 50] and upper bounds [0.99, 0.95, 3, 3, 1.2, 0.2, 150, 120] for the full parameter set $\theta=\{p_2,f,D_\mathrm{a},D_\mathrm{e}^{\parallel},D_\mathrm{e}^{\bot},f_\mathrm{w},T_2^\mathrm{a},T_2^\mathrm{e}\}$. In the context of two-shell dMRI protocols, the FW compartment is often omitted, as such protocols do not provide enough information to accurately retrieve parameters from all three compartments. To explore this limitation, we trained the SMI estimator under two conditions: excluding the FW compartment (2-shell woFW) and including it (2-shell wFW). In the 2-shell woFW scenario, the FW fraction was set to zero in all training samples. It is also important to note that fixed-TE and two-shell dMRI protocols are insensitive to $T_2$ relaxation times, as they are acquired for a single TE value. 

\subsection{Sensitivity-Specificity Matrix}
\noindent
Sensitivity and specificity are two key aspects for measuring the performance of model parameter estimation. In reality, the estimation of one parameter may be interfered by the others, creating spurious correlations and reducing specificity. To quantify sensitivity and spurious correlations of parameter estimation, we consider the 
{\it Sensitivity-Specificity Matrix} (SSM) in noise propagations 
\begin{equation}\label{eq_SSM}
S_{ij} = \frac{\langle\theta_i\rangle}{\langle\hat{\theta}_j\rangle}
\left \langle \frac{\partial \hat{\theta}_j}{\partial {\theta}_i}\right \rangle,\quad i,j=1,\hdots,N_{\theta} 
\end{equation}
whose elements quantify relative changes of an estimated parameter $\hat\theta_j$ with respect to the actual change of parameter $\theta_i$. Here, angular brackets denote averaging over the distribution of ground truths (the test set). The normalization by the mean values is introduced for convenience, to make the off-diagonal elements dimensionless. 

Practically, we evaluated the SSM from a linear regression of the estimates $\hat{\theta}_j$ with respect to ground truths ${\theta }_i$ of all $N_{\theta}$ parameters in a test set. The test set was sampled from a uniform distribution that covered the most probable range of the SM parameters $\theta$ 
 with lower bound [0.3, 0.3, 1.5, 1.5, 0.4, 0, 50, 40] and upper bound [0.8, 0.8, 2.5, 2.5, 1, 0.15, 120, 100]. In the simulation, the SNR was set at 40 for the $b_0$ measurements with a TE of 92 ms.
 
 Ideally, the SSM should be an identity matrix. The deviation from an identity matrix shows the limitations of the parameter estimation. In an SSM, the diagonal elements quantify the sensitivity, while the off-diagonal elements quantify spurious correlations and the lack of specificity. The elements of SSM are typically between -1 and 1, similar to the correlation coefficient, which makes them easier to interpret than RMSE.

\subsection{In vivo MRI data}
\noindent
After informed consent, six healthy volunteers, aged 24 to 60 years old (mean age: 33.8 $\mathrm{\pm}$ 13.4 years), including 3 females and 3 males, underwent brain MRI on a 3T Siemens Prisma, using a 64-channel head coil for reception. The total scan duration was 27 minutes, details in Fig. \ref{fig:fig1_protocol}. Due to the restricted availability of MRI sequences at the moment of scanning, we used the PTE shell with $\beta$ of $-0.5$, while the optimization resulted in $\beta$ value of 0.6. Maxwell-compensated asymmetric waveforms \citep{szczepankiewicz2019maxwell} were employed in the acquisition protocols using an in-house diffusion sequence that enables user defined gradient waveforms \citep{szczepankiewicz2019tensor}. A non-diffusion-weighted image with reverse phase-encoding was acquired to correct for EPI induced distortions \citep{andersson2003correct}. Imaging parameters: voxel size $2\times2\times2$ \unit{mm^3}, bandwidth 1818 Hz/pixel, GRAPPA factor 2, partial Fourier 6/8. Subjects were scanned, repositioned and subsequently rescanned with the same MRI protocol. The b-values and directions were randomized to avoid drift bias \citep{szczepankiewicz2021gradient}. 

Magnitude and phase data were reconstructed. This was followed by a process of phase estimation and unwrapping, which set the stage for denoising coil images \citep{lemberskiy2019achieving,veraart2016denoising}. After denoising, the coil images were combined and then processed by the DESIGNER pipeline \citep{ades2018evaluation,chen2023optimization}, including steps of denoising \citep{veraart2016denoising}, Gibbs ringing correction \citep{lee2021removal}, and correction for eddy current distortions and subject motion \citep{smith2004advances}. The last step was performed separately for each unique combination of ($\beta,\mathrm{TE}$). Finally, a rigid registration was computed to align all images to the subset with the largest number of DWI. 

Regions of interest (ROI) were automatically segmented by nonlinear mapping of the fractional anisotropy (FA) map onto the WM label atlas of Johns Hopkins University (JHU) \citep{mori2005}. FA maps for each subject were computed from a DKI fit based on the 2-shell subset of the data \citep{jensen2005diffusional}. Mean values of major WM regions were extracted after excluding outliers that fell outside of the physical range of SM parameters.

The full variable-TE protocol was used to jointly estimate the full SM parameter set $\theta$ for diffusion-relaxometry. A subset of this protocol acquired at $\mathrm{TE=92 ms}$ and a further subset with $b=0,\,1,\,2\,$\unit{ms/\mu m^2} were used for estimating SM parameters excluding $T_2$. The hierarchy of the variable-TE, fixed-TE and 2-shell dMRI protocol is illustrated in Fig. \ref{fig:fig1_protocol}, where the latter is a subset of the former. When comparing the variable-TE results with the fixed-TE and 2-shell dMRI protocols, we weighed the fractions $f$ and $f_\mathrm{w}$ estimated from the variable-TE protocol with corresponding $e^{-\mathrm{TE}/T_2}$ factors, and rescaled them to sum to 1. The rescaled fractions contained $T_2$ weighting and were denoted as $f^{\mathrm{(TE)}}$ and $f_{\mathrm{w}}^{\mathrm{(TE)}}$ for distinction. In addition to the above three protocols, we evaluated a clinically feasible 8-minute protocol by combining the 2-shell dMRI protocol (TE$=$92 ms) with $\mathrm{b_0}$ of multiple TEs (62, 78, 130 ms). This protocol, featuring varying TE, was specifically chosen to assess the feasibility of accurately estimating FW fractions using a protocol that is practical for clinical applications.

\subsection{Coefficients of variation} \label{cov}
\noindent
The scan and rescan of every subject were registered to one another for evaluating voxelwise reproducibility. Coefficients of variation (COV) were computed from the scan and rescan using the formula $\mathrm{COV}=\langle\sigma/\mu\rangle$, where the standard deviation was estimated as $\hat{\sigma}=\frac{\sqrt{\pi}}{2}\lvert x_1-x_2 \rvert$, and the mean $\mu$ was averaged over the scan and rescan of all available protocols. This ensures that the normalization for the same parameter across different protocols is consistent. The COV was first calculated for each voxel and then averaged over all the WM regions. We illustrated the COV for the six subjects in Fig. \ref{fig:fig4_cov}. The COV served as a reproducibility metric to assess the robustness of the acquisition protocol and parameter estimation. In addition, we employed the concordance correlation coefficient (CCC) to measure the consistency between two different protocols, and between the scan and rescan of the same protocol.

\section{Results}
\subsection{Protocol optimization}
\noindent
We incorporated several ``orthogonal" acquisition tactics, such as high b-values, multidimensional dMRI and multiple TE, into our optimized protocol by exploring the optimal combination of $(b, \beta, \mathrm{TE}, N_\mathrm{dirs})$. The tree structure in Fig. \ref{fig:fig1_protocol} demonstrates that the fixed-TE protocol is a subset of the variable-TE protocol, and the 2-shell dMRI protocol is a subset of the fixed-TE protocol. In the optimized variable-TE protocol, the fixed-TE segment from \cite{coelho2022reproducibility} was fixed into our optimization, while the remainder of the protocol was obtained directly as a result of our optimization framework. 

The 3D scatter plot of Fig. \ref{fig:fig1_protocol} illustrates the precise positioning of every shell in the acquisition space of $(b, \beta, \mathrm{TE})$. The 2-shell dMRI protocol includes three LTE shells at 0, 1, 2 \unit{\ms/\mu m^2}. Aside from that, the fixed-TE protocol consists of three shells: a high-b LTE shell at 6 \unit{\ms/\mu m^2}, a low-b PTE shell at 2 \unit{\ms/\mu m^2} and a low-b STE shell at 1.5 \unit{\ms/\mu m^2} \citep{coelho2022reproducibility}. In the fixed-TE protocol, all six shells are acquired at a TE of 92 ms. The variable-TE protocol extends this by including three additional TEs: 62, 78, and 130 ms. Within these additional shells of the variable-TE protocol, all feature a low b-value ($\leq$ 2 \unit{ms/\mu m^2}), with only one being a PTE shell and the others LTE shells. 

\begin{figure*}[htbp]
\centering
\includegraphics[width=0.9\textwidth]{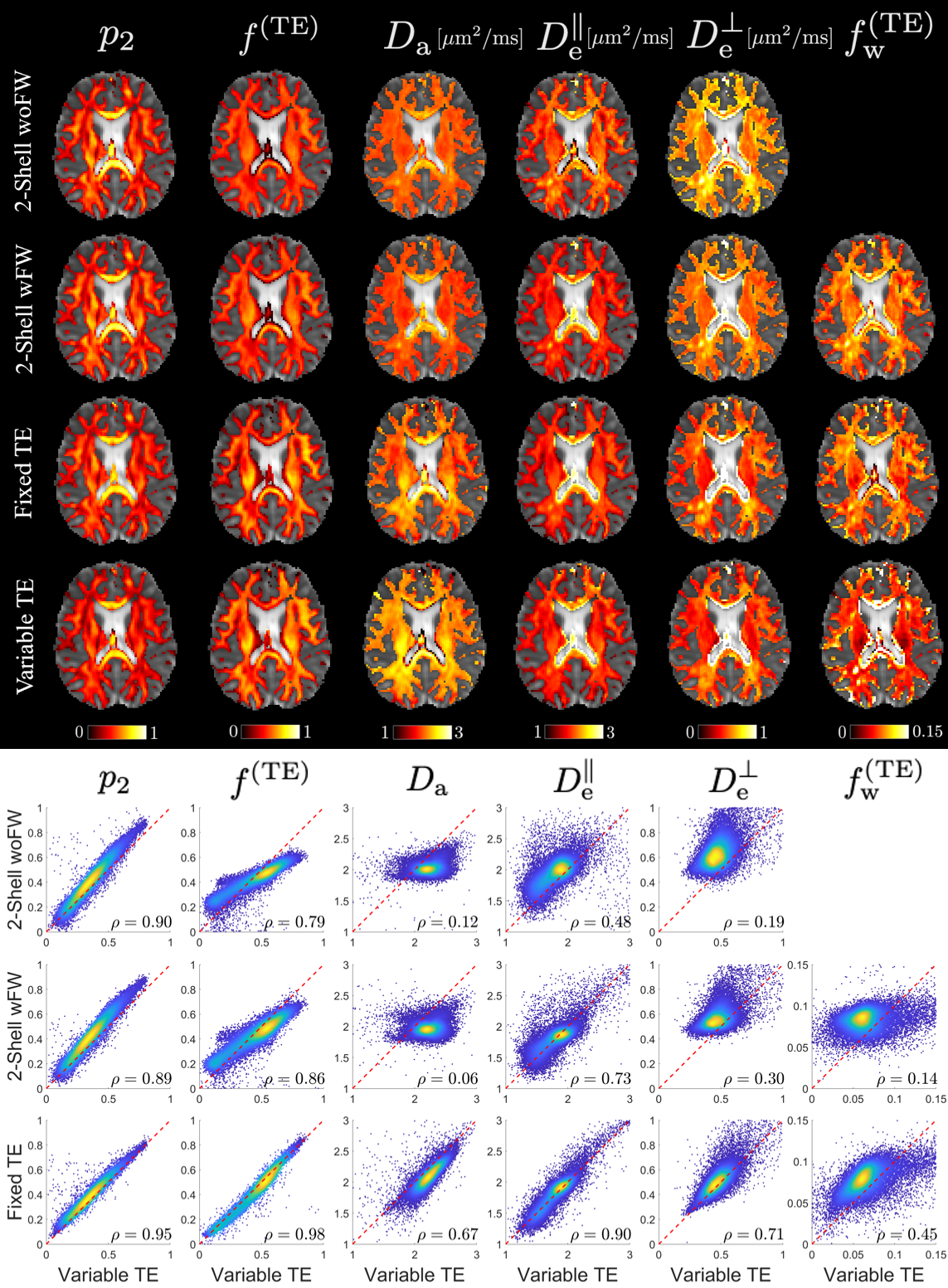}
\caption[caption FIG]{Parametric maps of SM parameters for the 3 protocols. Parametric maps of a 32-year-old male are shown for each protocol overlaid on the $b_0$ images. The mask for parametric maps is $\mathrm{FA}>0.2$. Scatter plots of voxelwise values from all WM ROIs are shown below with the concordance correlation coefficient $\rho$. $f^{(\mathrm{TE})}$ and $f^{(\mathrm{TE})}_{\mathrm{w}}$ are reweighed by estimated ${T}_2$ to compare with the ${T}_2$-weighted fractions in the 2-shell dMRI and fixed-TE protocols.}
\label{fig:fig3_3protocol}
\end{figure*}

\subsection{Sensitivity and specificity}
\noindent
Sensitivity and specificity are two key aspects of a biophysical model. Here, using the same ML parameter estimation algorithm and uniform prior distributions, we assessed the information content of the three protocols through their SSM (Fig. \ref{fig:fig2_ssm}). There are slight differences in the spurious correlations (off-diagonal elements of SSM) between 2-shell woFW and 2-shell wFW. Specifically, when the FW compartment is included in the training data, the correlation between the estimated ${D}_\mathrm{e}^{\bot}$ and the ground truth of $f$ increases. Conversely, when FW is excluded, the estimated ${D}_\mathrm{e}^{\bot}$ shows a higher correlation with the ground truth of $f_{\mathrm{w}}$. 

All parameters improve with additional ``orthogonal" acquisitions, though this comes at the cost of increased acquisition time. For $f$ and $p_2$, which are very accurate already with 2-shell, the spurious correlations between them and compartmental diffusivities decrease by adding free gradient waveforms (fixed-TE) and get further reduced by adding multiple TE (variable-TE). $D_\mathrm{a}$ and $D_\mathrm{e}^{\bot}$ estimations are substantially improved in fixed-TE and even more so in variable-TE. In particular, $D_\mathrm{a}$ reaches a performance close to $p_2$ and $f$ with variable-TE. On the contrary, $D_\mathrm{e}^{\parallel}$ shows the least improvement of accuracy after adding multidimensional dMRI and varying TE, indicating it is the least sensitive parameter.

The obvious gain from including varying TE is the ability to capture $T_{2}^{\mathrm{a}}$ and $T_{2}^{\mathrm{e}}$, to which 2-shell and fixed-TE protocols are insensitive. The estimation of $T_{2}^{\mathrm{a}}$ is at the level of $p_2$ and $f$, and therefore very good, with the estimation of $T_{2}^{\mathrm{e}}$ showing a lower but still reasonable sensitivity. Yet, the most remarkable improvement happens to $f_{\mathrm{w}}$. In the 2-shell dMRI protocol, estimating the FW fraction is unfeasible. After including multidimensional dMRI, the estimated ${f}_{\mathrm{w}}$ still exhibits strong spurious correlations with $D_\mathrm{e}^{\parallel}$ and $D_\mathrm{e}^{\bot}$. In contrast, the variable-TE protocol demonstrates high sensitivity in estimating $f_{\mathrm{w}}$ within a range of 0 to 0.15 in the test set, with the sensitivity metric at 0.99 (diagonal elements of SSM). The improvement for $f_{\mathrm{w}}$ is expected since the signal contrast between FW and the remaining compartments increases when varying TE.

\begin{figure*}[b]
\centering
\includegraphics[width=\textwidth]{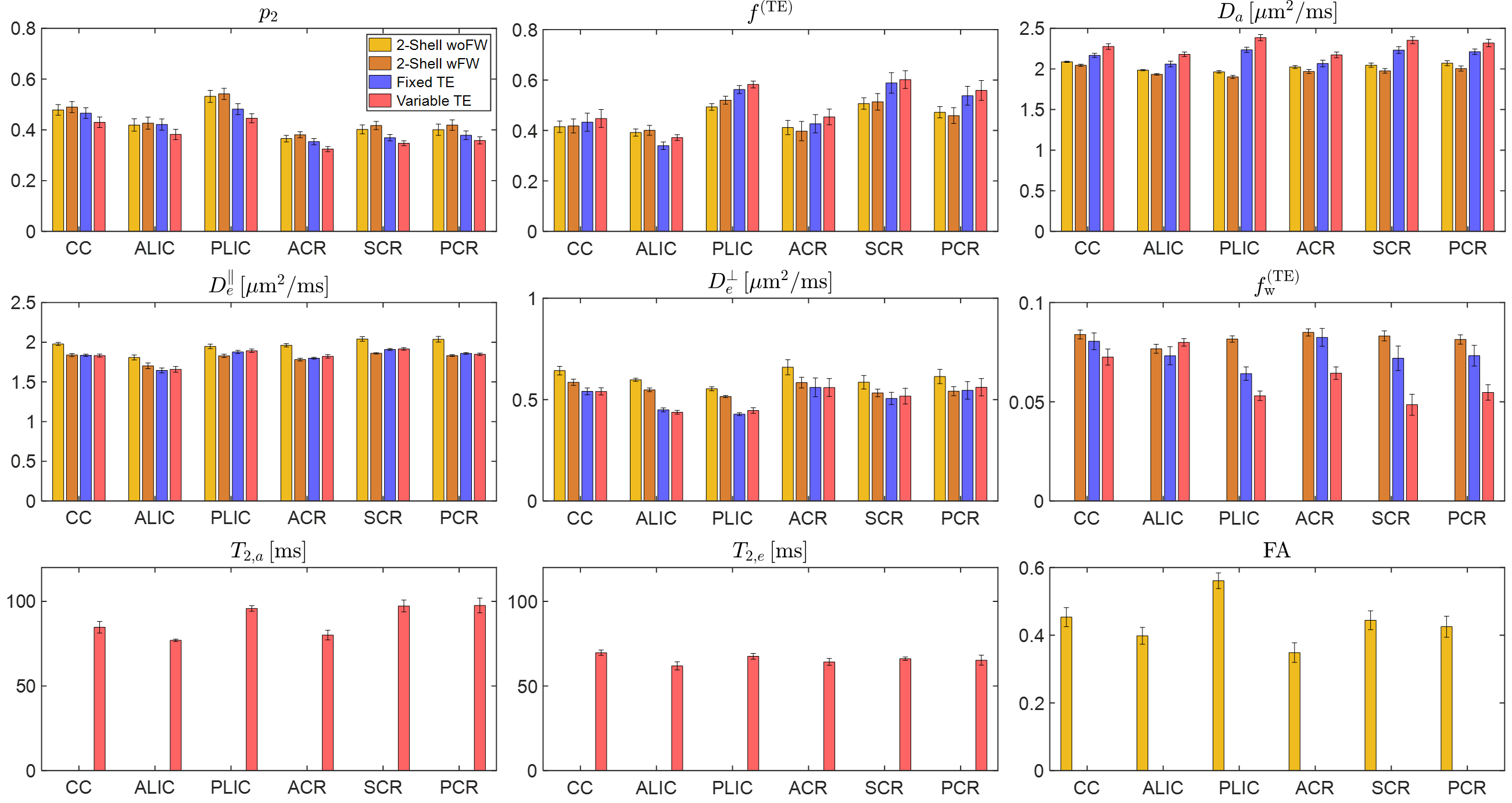}
\caption[caption FIG]{ROI means of SM parameters for the 3 protocols. Mean values of major WM regions from 6 subjects are shown for different protocols. $\mathrm{T_{2}}$ relaxation times are shown for the variable TE protocol alone, and FA values are calculated using only the 2-Shell protocol. Here, fractions $f^{\mathrm{(TE)}}$ and $f_{\mathrm{w}}^{\mathrm{(TE)}}$ are $\mathrm{T_{2}}$-weighted, hence the notation TE in superscript. The error bar of the bar plots indicates the standard deviation of mean ROI values among the 6 subjects. ROI abbreviations: CC, corpus callosum; ALIC, anterior limb of internal capsule; PLIC, posterior limb of internal capsule; ACR, anterior corona radiata; SCR, superior corona radiata; PCR, posterior corona radiata.}
\label{fig:fig_roi_mean}
\end{figure*}

\subsection{Parametric maps}
\noindent
The parametric maps of the three protocols and their voxelwise scatter plots are shown in Fig. \ref{fig:fig3_3protocol}. The parametric maps of $f$ and $D_a$  show weaker contrast variation for the 2-shell protocol as compared to the other two richer protocols. In the scatter plots, we use the CCC ($\rho$) as the consistency metric to take into account the difference in the mean and standard deviation between two sets of estimations. Using estimates from variable-TE as the ground truth, as expected, fixed-TE has higher accuracy than 2-shell dMRI, and $p_2$ and $f$ are overall more accurate than compartment diffusivities. Interestingly, 2-shell wFW has slightly higher consistency with variable-TE for $f$, $D_\mathrm{e}^{\parallel}$ and $D_\mathrm{e}^{\bot}$ than 2-shell woFW. Moreover, the disparity between variable-TE and its subsets highlights the extent of improvement gained through additional measurements. For instance, in the comparison between 2-shell wFW and variable-TE,  $D_\mathrm{a}$ ($\rho=0.06$) and $f_{\mathrm{w}}^{\mathrm{(TE)}}$ ($\rho=0.14$) show the lowest correlations and thus the largest improvement after including additional data, followed by $D_\mathrm{e}^{\bot}$ ($\rho=0.30$). When comparing fixed-TE with variable-TE, $f_{\mathrm{w}}^{\mathrm{(TE)}}$ sees the largest improvement after adding shells with varying TE. Yet, the consistency levels between 2-shell wFW, fixed-TE and variable-TE are relatively high for $D_\mathrm{e}^{\parallel}$, suggesting minimal improvement for this parameter after acquiring more data. These findings are consistent with our previous analysis of the SSM. 

Fig. \ref{fig:fig_roi_mean} displays the ROI means of six major WM regions for the three protocols. Generally, the ROI mean differences between these protocols align with the voxelwise biases illustrated in the scatter plot of Fig. \ref{fig:fig3_3protocol}. The more advanced variable-TE protocol often yields estimates that deviate further from the mean of prior distributions, highlighting that the estimates are more informed by actual measurements rather than being dominated by the prior distribution. For instance, considering the prior mean for $D_{\mathrm{a}}$ at 2 \unit{\mu m^2/ms}, variable-TE shows $D_{\mathrm{a}}$ approximately 2.4 \unit{\mu m^2/ms} in the posterior limb of internal capsule (PLIC), while the 2-shell protocol yields $D_{\mathrm{a}}$ close to the prior mean and fixed-TE shows intermediate values. Furthermore, the variable-TE protocol exhibits greater biological variability across different ROIs, while the 2-shell protocol may yield nearly uniform values for all ROIs, a trend that is particularly pronounced in the case of $f_{\mathrm{w}}$. In addition, using variable-TE as a reference, 2-shell wFW demonstrates smaller biases than 2-shell woFW for $D_\mathrm{e}^{\bot}$ and $D_\mathrm{e}^{\parallel}$.

\begin{figure*}[t]
\centering
\includegraphics[width=\textwidth]{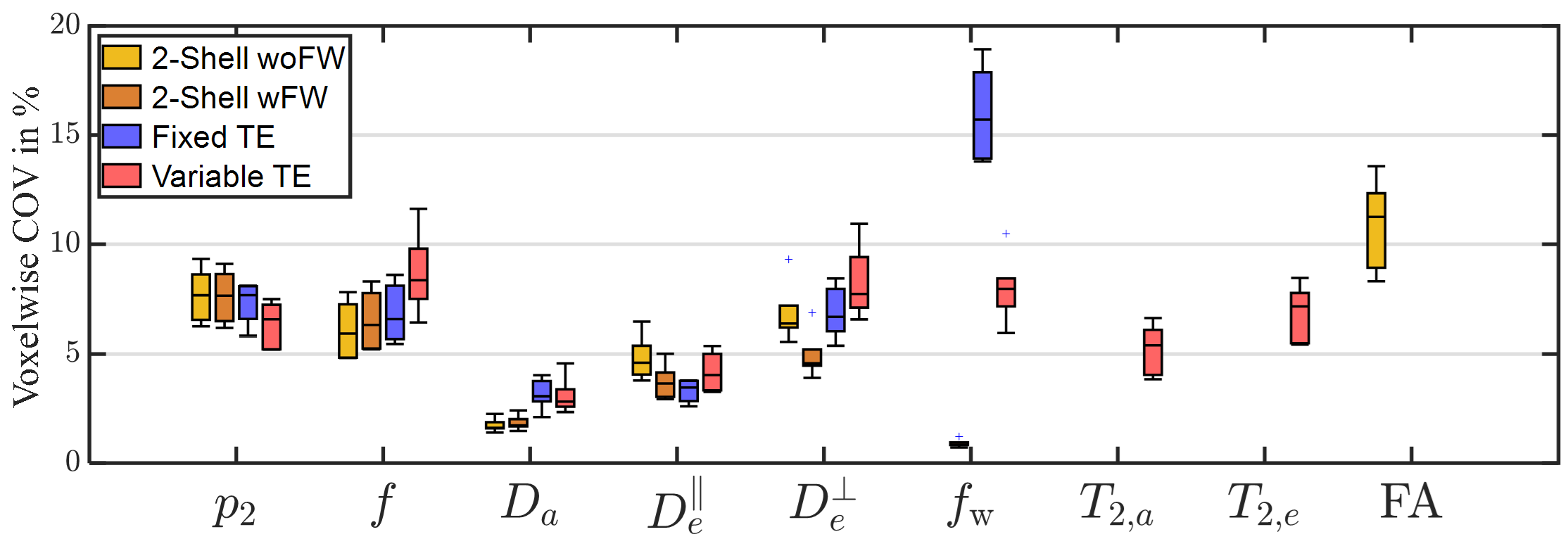}
\caption[caption FIG]{Reproducibility of SM parameter estimation. The coefficient of variation (COV) is mostly below 10\% for almost all SM parameter. The COV of FW fractions for 2-shell is extremely low due to the dominance of the prior distribution in parameter estimation, while the COV of fixed-TE is notably high because of the small values of FW fractions in the normalization of COV. Note that fractions for the 2-shell dMRI and fixed-TE protocol are $\mathrm{T_2}$-weighted, and fractions for the variable-TE protocol are not. We use the same symbol for every protocol for clarity.}
\label{fig:fig4_cov}
\end{figure*}

\subsection{Reproducibility}
\noindent
Fig. \ref{fig:fig4_cov} plots the voxelwise COV of each SM parameter based on a scan and rescan using the three protocols. Overall, the COV of most SM parameters are below 10\%. The COV of 2-shell for $f_{\mathrm{w}}$ is extremely low, and the variable-TE protocol significantly reduces the variation of $f_{\mathrm{w}}$ between scan and rescan compared to fixed-TE.

Fig. \ref{fig:fig5_scan_rescan} presents the parametric maps of the scan and rescan using the variable-TE protocol. FA maps are plotted alongside SM parameters for reference. The CCC is calculated for all WM ROIs to evaluate the consistency between the scan and rescan. Parameters $p_2$ ($\rho=0.97$) and $f$ ($\rho=0.95$) demonstrate the highest reproducibility, comparable to that of FA ($\rho=0.96$). The reproducibility of all parameters in the variable-TE protocol are reasonably good, with the lowest CCC being above 0.7.

\subsection{Free water estimation}
\noindent
Free water estimation is challenging with only 2-shell protocols, and is still hampered by spurious correlations with EAS diffusivities using the fixed-TE protocol (Fig. \ref{fig:fig2_ssm}). However, employing a variable-TE protocol considerably enhances the sensitivity to $f_{\mathrm{w}}$, prompting us to investigate the effectiveness of a clinically feasible protocol that combines a 2-shell dMRI protocol with $\mathrm{b_0}$ measurements at three distinct TEs (62, 78, 130 ms) for FW estimation. Remarkably, this 8-minute protocol achieves sensitivity of 0.87 for $f_{\mathrm{w}}$ and maintains spurious correlations below 0.30 (Fig. \ref{fig:fig_fw}A). Although resolving other parameters like compartmental diffusivities and compartmental $T_2$ remains challenging, this 8-minute protocol, varying TE solely for $\mathrm{b_0}$ measurements, is sensitive to FW fractions. Fig. \ref{fig:fig_fw}B shows a scatter plot comparing $f_{\mathrm{w}}$ estimates obtained from the streamlined 8-minute protocol against those from a more comprehensive 27-minute protocol. The CCC is 0.80 between these two protocols for $f_{\mathrm{w}}$, exceeding the CCC of 0.45 between fixed-TE and variable-TE shown in Fig. \ref{fig:fig3_3protocol}. This result highlights the benefit of varying TE for estimating free water fractions.

\section{Discussion}
\noindent
In this paper, we presented an optimized 27-minute MRI acquisition protocol that extends conventional 2-shell protocols with three major features: high b-value, multiple B-tensor shapes and multiple TE. This protocol enables a simultaneous measurement of both diffusion and relaxometry. We found the optimal protocol in the sampling space of $(b,\beta,\mathrm{TE})$ by minimizing the RMSE of an ML-based estimator (SMI). This is different from the work from \cite{lampinen2020towards} where CRLB was the objective function and MLE was adopted to estimate parameters. Our objective function is tailored for the ML-based estimation, as the analytical solution of RMSE takes into account the distribution of training sets  \citep{coelho2021we}. Then, we evaluated the sensitivity, specificity and reproducibility of the proposed 27-minute protocol and its subsets, a 7-minute 2-shell dMRI protocol and a 15-minute fixed-TE protocol, to demonstrate the gain in each model parameter from adding extra measurements.

\subsection{Protocol optimization}
\noindent
In the optimization process, we integrated the fixed-TE protocol \citep{coelho2022reproducibility} into our optimized design to enable comparisons. The final optimized protocol, which already includes multidimensional dMRI and high-b shells, was further enhanced by adding shells with variable TE. These added shells feature a TE range from 62 ms to 130 ms, compared to the uniform TE of 92 ms in the fixed-TE protocol. All of these additional shells have b-value between 0 and 2 \unit{\ms/\mu m^2}, with 7 out of 8 additional shells being LTE. This configuration suggests that the variable-TE shells are primarily used to assess signal decay with respect to TE in the low-b regime for LTE. 

\subsection{Sensitivity and specificity}
\noindent
Sensitivity and specificity of the different MRI protocols were quantified by the SSM. The SSM is conceptually similar to the confusion matrix in classification problems. In a classification problem, the prediction of a certain category may contain misclassified samples from other categories. Likewise, in a regression problem, the prediction of a certain parameter may contain information from other parameters. The spurious correlations undermine the core motivation behind biophysical modeling -- specificity. Despite the effort to parameterize a quantity directly related to microstructure, estimating a nonlinear model parameter from limited noisy data can result in biases and spurious correlations. Constraints on the SM may yield relatively more precise results \citep{zhang2012,kaden2016,fieremans2011}, but they do not resolve the issue of spurious correlations between model parameters \citep{liao2023mapping}. We demonstrate that acquiring ``orthogonal" data is essential to address these issues effectively.

\begin{figure*}[htbp]
\centering
\includegraphics[width=\textwidth]{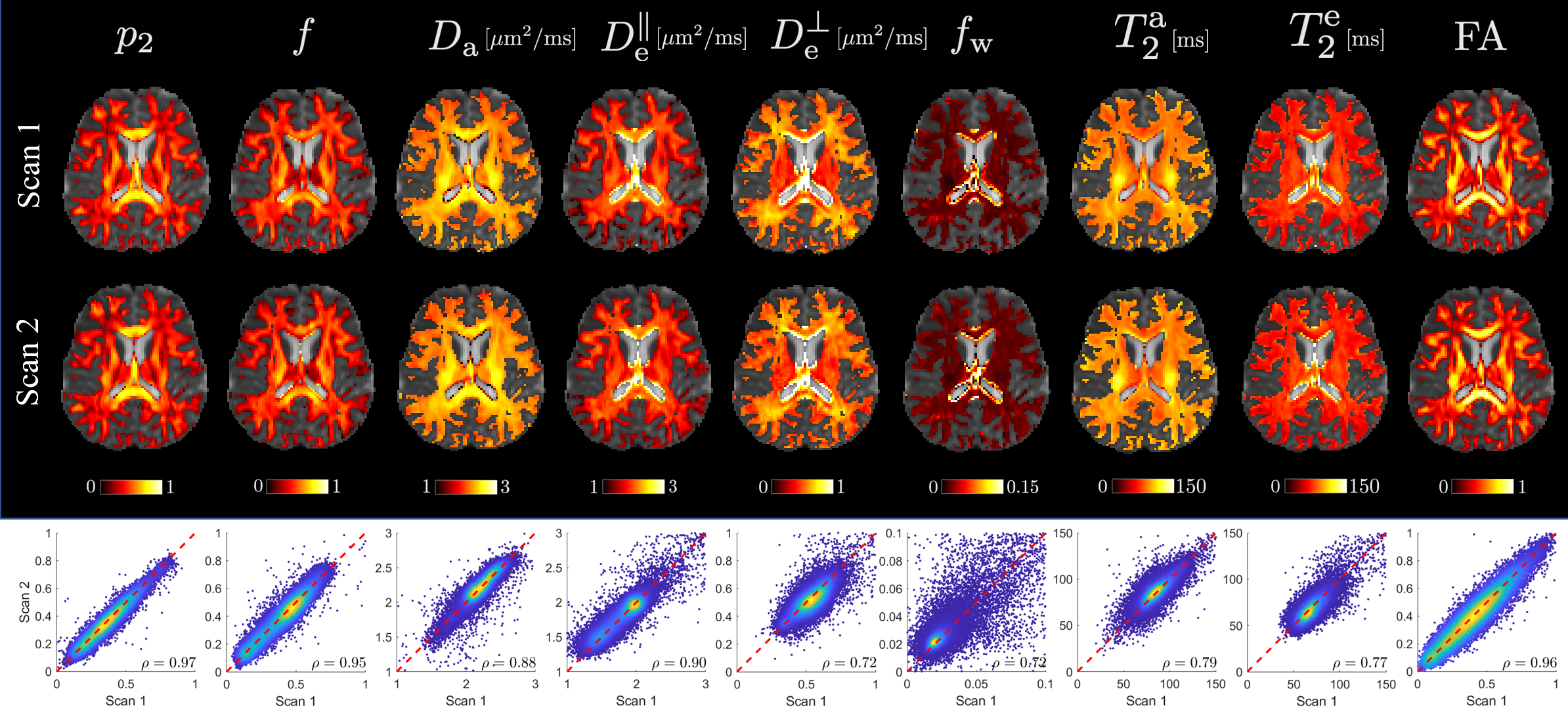}
\caption[caption FIG]{Parametric maps of SM parameters for the variable-TE protocol. Parametric maps of a 28 year-old male are shown for the scan and rescan of variable-TE protocol overlaid on $b_0$ images. The maps are masked by $\mathrm{FA}>0.2$. Scatter plots of voxelwise values from all WM ROIs are shown below with the concordance correlation coefficient $\rho$.}
\label{fig:fig5_scan_rescan}
\end{figure*}

The 2-shell dMRI protocol of this study is largely the same as protocols adopted in some large imaging consortia, such as UK Biobank \citep{miller2016}, Human Connectome Project \citep{glasser2016}, Alzheimer's Disease Neuroimaging Initiative \citep{jack2008} and Adolescent Brain Cognitive Development \citep{casey2018}. Here, we study the gain of incorporating additional measurements by comparing 2-shell dMRI and fixed-TE with variable-TE protocols. With high b-value and multiple B-tensor shapes in the fixed-TE protocol, estimates of $D_\mathrm{a}$ and $D_\mathrm{e}^{\bot}$ become much more sensitive and specific, while estimating $f_{\mathrm{w}}$ and $D_\mathrm{e}^{\parallel}$ remain more challenging. Furthermore, varying TE improves the estimation of $f_{\mathrm{w}}$, which is anticipated due to the substantial difference between $T_2$ of FW and that of the IAS or EAS. The variable-TE protocol also reliably estimates $T_{2}^{\mathrm{a}}$ and $T_{2}^{\mathrm{e}}$, both of which have the potential to become biomarkers for WM pathology. $D_\mathrm{e}^{\parallel}$ is improved in variable-TE, but caution is advised for its interpretation due to the relatively low sensitivity and specificity.

\begin{figure*}[t]
\centering
\includegraphics[width=\textwidth]{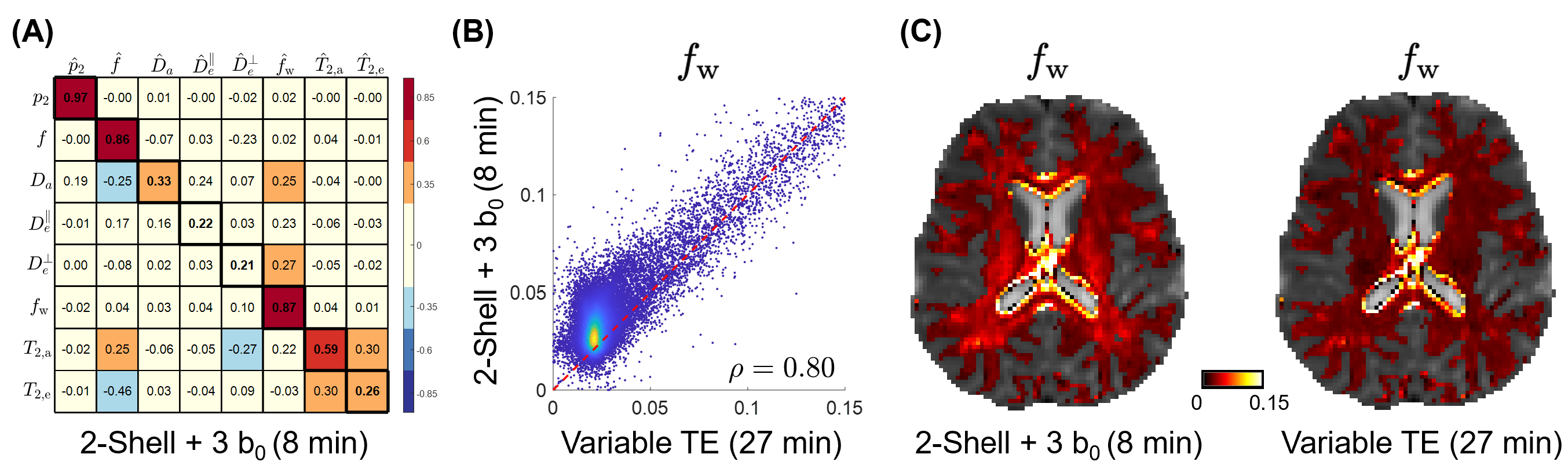}
\caption[caption FIG]{Free water estimation using the 2-shell protocol augmented with $\mathrm{b_0}$ of varying TE. Panel A shows the SSM for a clinically-feasible 8-minute MRI protocol, which integrates the 2-shell dMRI protocol (TE$=$92 ms) with $\mathrm{b_0}$ measurements at varying TE (62, 78, 130 ms), specifically designed to estimate FW fractions. Panel B presents a scatter plot comparing $f_{\mathrm{w}}$ estimated from this 8-minute protocol against those from a more extensive 27-minute variable-TE protocol, utilizing data from the same 32-year-old subject previously shown in Fig. \ref{fig:fig3_3protocol}. The concordance correlation coefficient $\rho$ between these two protocols is indicated on the scatter plot. Please note that, here, $f_{\mathrm{w}}$ is not $T_2$-weighted, different from $f_{\mathrm{w}}^{\mathrm{(TE)}}$ shown in Fig. \ref{fig:fig3_3protocol}. Panel C displays $f_{\mathrm{w}}$ parametric maps of this subject generated by both protocols, illustrating the spatial distribution of FW content across the WM.}
\label{fig:fig_fw}
\end{figure*}

To estimate FW fractions for clinical applications, We designed a clinically feasible 8-minute protocol by combining the 2-shell dMRI protocol with $\mathrm{b_0}$ measurements at three different TEs. This protocol does not require free gradient waveforms, and only extends the standard 2-shell protocol by adding several $\mathrm{b_0}$ measurements of multiple TEs. This protocol, a small subset of the 27-minute variable-TE protocol, demonstrates remarkably high sensitivity to $f_{\mathrm{w}}$ (Fig. \ref{fig:fig_fw}). It outperforms the 15-minute fixed-TE protocol for $f_{\mathrm{w}}$ in terms of the SSM as well as the CCC when comparing against the 27-minute variable-TE protocol. Fig. \ref{fig:fig_fw}B shows that the CCC is 0.80 for the entire range of $f_{\mathrm{w}}$, with even greater consistency observed for free water fractions over 5\%, compared to the CCC of 0.45 for fixed-TE protocols against variable-TE for $f_{\mathrm{w}}$ in Fig. \ref{fig:fig3_3protocol}. This indicates that varying TE can enhance $f_{\mathrm{w}}$ estimation more effectively than using multidimensional dMRI. Nevertheless, this protocol falls short in accurately resolving compartmental diffusivities and $T_2$ values. When obtaining $f_{\mathrm{w}}$ is of interest in clinical settings, a standard 2-shell dMRI protocol could be augmented with an additional minute of scanning time to include $\mathrm{b_0}$ measurements at multiple TEs.

\subsection{Reproducibility}
\noindent
We assessed the reproducibility of three protocols by scan/rescanning of six subjects. Overall, the COV of most parameters is between 5--10\% voxelwise, which is comparable to FA. The COV of $f_{\mathrm{w}}$ for fixed-TE protocols is high, in part due to the normalization by their small values as denominators. Yet, the reproducibility improved with variable-TE protocols, highlighting the benefit of incorporating multiple TE to estimate $f_{\mathrm{w}}$. Surprisingly, the 2-shell dMRI protocol exhibits an extremely low COV for $f_{\mathrm{w}}$, as well as for some parameters that are typically difficult to estimate, such as $D_a$. This somewhat unphysically low variation stems from the protocol’s estimates being heavily influenced by the prior distribution. In light of that, the COV for ML-based estimators should be interpreted with caution and, ideally, in conjunction with the SSM. 

The CCC is identified as an alternative metric for assessing reproducibility. In contrast to COV, which is calculated on a voxel-by-voxel basis before averaging over an ROI, CCC evaluates the entire dataset collectively. This characteristic makes CCC less susceptible to extreme values or small denominators in the normalization process. Besides, CCC accounts for the difference in the mean and standard deviation of two sets of measurements, going beyond just the Pearson correlation. As a standard metric for consistency, CCC is naturally suited for measuring reproducibility. Results from Fig. \ref{fig:fig5_scan_rescan} suggest that $p_2$ and $f$ are the only two parameters demonstrating a level of reproducibility comparable FA. This finding aligns with their estimation performance shown in the SSM (Fig. \ref{fig:fig2_ssm}).

\subsection{Training data}
\noindent
The selection of training sets, which serve as the prior distribution in this Bayesian approach, is crucial in regularizing ML-based estimators. These estimators are designed to minimize the overall RMSE across all training samples, rather than optimizing for each individual sample. In doing so, these estimators tend to strike a balance between bias and precision. For parameters that are more challenging to estimate, the estimator often shifts predictions toward the prior mean to enhance precision, albeit at the expense of increased bias. This approach aligns with Bayesian principles, where parameter predictions are based on a pre-existing belief about their distribution \citep{reisert2017}. When estimating difficult parameters, adopting a more informative and narrow prior can enhance performance within the most probable parameter range \citep{liao2023mapping}. However, the inherent dependency of ML-based estimators on the prior distribution may introduce a systematic bias to the overall estimation if the SNR is too low or measurements are insufficient.  Note that this effect is present in any data-driven estimator minimizing RMSE and is mediated by the SNR and the measurements acquired. Thus, to ensure consistency across different estimations, the same prior distribution needs to be used to train the ML-based estimator.

The bias due to the prior distribution is evident in the relatively flat parametric maps of the 2-shell dMRI protocol shown in Fig. \ref{fig:fig3_3protocol}. Specifically, for parameters like $f_{\mathrm{w}}$, which are challenging to estimate at low values using only the 2-shell protocol, the sensitivity and specificity are the lowest among all parameters. As a result, the ROI mean values of $f_{\mathrm{w}}$ for the 2-shell protocol appear uniformly close to the prior mean. The lack of variation in $f_{\mathrm{w}}$ also leads to an extremely low COV for 2-shell protocols between the scan and rescan. However, knowing that the FW fraction typically ranges between 5 and 10\% from the parametric maps of variable-TE protocols, incorporating a similar amount of FW in the training samples for 2-shell protocols can mitigate bias in EAS compartment diffusivities when comparing 2-shell wFW with 2-shell woFW (Fig. \ref{fig:fig_roi_mean}). 

As more ``orthogonal" data is acquired, the dependence of estimates on the prior distribution decreases. ROI mean values from the variable-TE protocol tend to diverge more from the prior mean than those from less comprehensive protocols. While the 2-shell protocol shows both $D_{\mathrm{a}}$ and $D_{\mathrm{e}}^{\parallel}$ to be around 2 \unit{\mu m^2/ms} (prior mean), the variable-TE protocol reveals that $D_{\mathrm{a}}$ is consistently higher than $D_{\mathrm{e}}^{\parallel}$ across all examined ROIs. In addition, the variable-TE protocol captures a greater biological variability, as reflected in both the ROI means and the parametric maps. To enhance the sensitivity and specificity of SM parameter estimation, it is essential to incorporate additional measurements that can provide complementary information about the underlying tissue microstructure.

\subsection{Parameter estimates in vivo}
\noindent
In addition to the simulation showing excellent performance (Fig. \ref{fig:fig2_ssm}), the parameter estimates by the variable-TE protocol (Fig. \ref{fig:fig5_scan_rescan}) generally agree with the prior studies. The fraction of axons $f$ shows the highest values in the most densely packed regions, such as corpus callosum and internal capsule. The mean value of IAS axial diffusivity $D_\mathrm{a}$ is about 2.2 \unit{\mu m^2/ms} in WM, consistent with the result from planar signal filtering \citep{dhital2019intra}. The FW fraction $f_{\mathrm{w}}$ is very high near ventricles due to the partial volume effect of cerebrospinal fluid and fluid around the brain parenchyma. Otherwise, $f_{\mathrm{w}}$ is between 0 and 5\% in most WM voxels, and the $T_2$- weighted FW fractions $f_{\mathrm{w}}^{\mathrm{(TE)}}$ generally fall between 5 and 10\%. $T_{2}^{\mathrm{a}}$ of the IAS is higher near ventricles also due to partial volume effect. $T_{2}$ ranges from 60 to 120 ms in the IAS and is generally higher than that of the EAS, which is consistent with findings in the human brain \citep{mckinnon2019measuring,veraart2018te} and in ex vivo animal nervous tissues \citep{peled1999water,wachowicz2002assignment,bonilla2007transverse,dortch2010compartment}. 

\subsection{Limitations and outlook}
\noindent
In this section, we summarize the limitations and offer future outlook for this study. 
First, myelin water is characterized by a particularly short $T_2$ relaxation time \citep{mackay1994vivo} compared to the other compartments of WM, as its signal decays rapidly after the MRI pulse. Due to the diffusion gradients employed in our MRI protocols, the TE of our measurements is beyond the range for capturing myelin water fractions. However, we expect the presence/absence of myelin to impact $D_\mathrm{e}^\bot$ \citep{jelescu2016vivo}. $D_\mathrm{e}^\bot$ can serve as a proxy for  myelin water fractions. 
Second, when comparing the three protocols, different models were employed. There are no $T_2$ relaxation parameters in the 2-shell dMRI and fixed-TE protocol, as they are acquired at only one TE. However, after reweighing fractions with their corresponding compartmental $T_2$, $f^{\mathrm{(TE)}}$ and $f_{\mathrm{w}}^{\mathrm{(TE)}}$ from the variable-TE protocol achieve excellent agreement with those estimated from fixed-TE (Fig. \ref{fig:fig3_3protocol}). This result provides an important consistency check for SM assumptions, as the fractions and $T_2$ relaxation times can maintain their relationship when estimated separately. 
Third, we did not explore the optimization landscape or the protocol parameter in detail in this manuscript. Readers can refer to \cite{coelho2022reproducibility} for more discussions. The optimization landscape as a function of protocol parameters is quite flat around this protocol, and that means small changes in the shell settings will hardly affect the ultimate performance. 
Fourth, the time dependence of diffusion is ignored in this study, as we assume that for our experimental settings diffusion in WM has reached the long-time limit and thus every compartment can be modeled as Gaussian. 
Fifth, even in the variable-TE protocol, $D_\mathrm{e}^{\parallel}$ still has relatively low sensitivity and needs to be improved further for obtaining a robust estimation of the EAS axial diffusivity.
Sixth, the 27-minute variable-TE protocol is time-efficient for estimating the entire set of SM parameters, as the optimization framework was aimed for. However, to obtain a specific parameter, a less comprehensive protocol that is practical for clinical settings might be adequate. The 8-minute protocol extending 2-shell protocols with $\mathrm{b_0}$ of multiple TEs can achieve fairly high sensitivity to $f_{\mathrm{w}}$. Future studies can focus on optimizing the protocol for a selection of SM parameters, thereby addressing specific clinical needs. 

\section{Conclusion}
\noindent
In this work, we presented a protocol optimization framework that generated a 27-minute variable-TE protocol for joint modeling of diffusion-relaxation. The variable-TE protocol was compared to its subsets, a fixed-TE protocol and a 2-shell dMRI protocol. The results show that by varying TE, FW fractions can now be accurately estimated even at very low values, which is challenging with the fixed-TE protocol. In addition, compartment $T_2$ values are now simultaneously estimated with high accuracy. In terms of reproducibility for variable-TE protocols, COV are below 10\% for almost all SM parameters, comparable to FA. Our work demonstrates the benefits of multidimensional dMRI and varying TE to the parameter estimation. The variable-TE protocol can become a powerful tool for studying brain tissue microstructure.

\section*{Acknowledgments}
\noindent
This work was performed under the rubric of the Center for Advanced Imaging Innovation and Research (CAI2R, \href{https://www.cai2r.net}{https://www.cai2r.net}), a NIBIB Biomedical Technology Resource Center (NIH P41-EB017183). This work has been supported by NIH under NINDS award R01 NS088040 and NIBIB awards R01 EB027075. This work was also supported by the Swedish Cancer Society (22 0592 JIA).

\section*{Author contributions}
\noindent
Conceptualization: S.Co, Y.L., J.V., D.S.N., E.F.; Data curation: S.Co, Y.L.; Formal analysis: S.Co, Y.L.; Funding: Y.W.L., D.S.N., E.F.; Investigation: S.Co, Y.L., F.S., J.V., S.Ch., Y.W.L., D.S.N., E.F.; Visualization: S.Co, Y.L.; Software: S.Co, Y.L, F.S.; Project administration: D.S.N, E.F.; Supervision: D.S.N., E.F.; Writing - original draft: S.Co, Y.L.; Writing - review \& editing: S.Co, Y.L., F.S., J.V., S.Ch., Y.W.L., D.S.N., E.F..

\bibliographystyle{elsarticle-harv}
\bibliography{bibliography.bib}

\end{multicols}

\end{document}